\documentclass[journal]{IEEEtran}

\usepackage{graphicx}
\usepackage{subcaption}

\usepackage{amsmath}

\usepackage{url}

\begin{document}
%
\title{Analysis of Altitude-Dependent Electronic Conspicuity in Cellular-Connected UAVs
}
\author{
Md~Sharif~Hossen,~\IEEEmembership{Senior~Member,~IEEE,}
Vijay~K.~Shah,~\IEEEmembership{Senior Member,~IEEE,}
and~Ismail~Guvenc,~\IEEEmembership{~Fellow,~IEEE}\\
\IEEEauthorblockA{Department of Electrical and Computer Engineering\\
North Carolina State University, Raleigh, NC, USA\\
Email: \{mhossen, vijay.shah, iguvenc\}@ncsu.edu
}
\thanks{This work is supported in part by the NASA ULI award 80NSSC25M7102.}
}

\markboth{Journal of \LaTeX\ Class Files,~Vol.~14, No.~8, August~2015}%
{Shell \MakeLowercase{\textit{et al.}}: Bare Demo of IEEEtran.cls for IEEE Journals}

\maketitle
\begin{abstract}
Unmanned aerial vehicles (UAVs) are increasingly integrated into cellular networks to support emerging Internet of Things (IoT) applications. In such settings, reliable communication is critical for electronic conspicuity (EC), enabling UAV detection and tracking in shared airspace. However, UAVs operate at elevated altitudes where enhanced line-of-sight (LOS) visibility leads to simultaneous exposure to multiple base stations, resulting in strong inter-cell interference. This article presents a system-level analysis of how UAV altitude influences the radio environment and affects EC reliability. Using spatial and network-level metrics, including serving distance, association behavior, and aggregate received power, we show that increasing altitude leads to stronger multi-cell interaction, reduced dominance of nearby sectors, and interference-dominated connectivity. These effects result in fragmented association regions and increased variability in link performance. The analysis is supported by measurement data from a helikite-based spectrum monitoring campaign and corresponding simulation results. Despite differences in experimental conditions, both approaches exhibit consistent altitude-dependent trends. These findings provide practical insights for designing altitude-aware and interference-aware cellular systems to support reliable UAV operation.

\end{abstract}
\IEEEpeerreviewmaketitle

\section*{Introduction}
\label{sec:introduction}
\IEEEPARstart{T}{he} rapid growth of Internet of Things (IoT) applications is driving an unprecedented demand for reliable and scalable wireless connectivity. Emerging use cases such as aerial sensing, data collection, infrastructure inspection, and emergency response increasingly rely on unmanned aerial vehicles (UAVs) to extend coverage and provide flexible network access~\cite{Granelli11435927}. In many of these applications, continuous and reliable connectivity is essential not only for data exchange but also for electronic conspicuity (EC), i.e., the ability of UAVs to be detected and tracked by other airspace users. Ensuring EC requires consistent communication performance, which directly depends on the underlying radio environment experienced by the UAV~\cite{EC_nassif2024}. Existing aviation safety systems, such as automatic dependent surveillance-broadcast (ADS-B) and the traffic collision avoidance system (TCAS), provide aircraft surveillance and collision-avoidance support. However, these systems were primarily designed for conventional aviation and may not fully address the scalability, altitude-dependent connectivity, and spectrum coexistence challenges introduced by dense UAV operations.

\begin{figure}[t]
    \centering
    \includegraphics[width=1.2\linewidth,trim=2.2in 2.1in 1in 1.8in,clip]{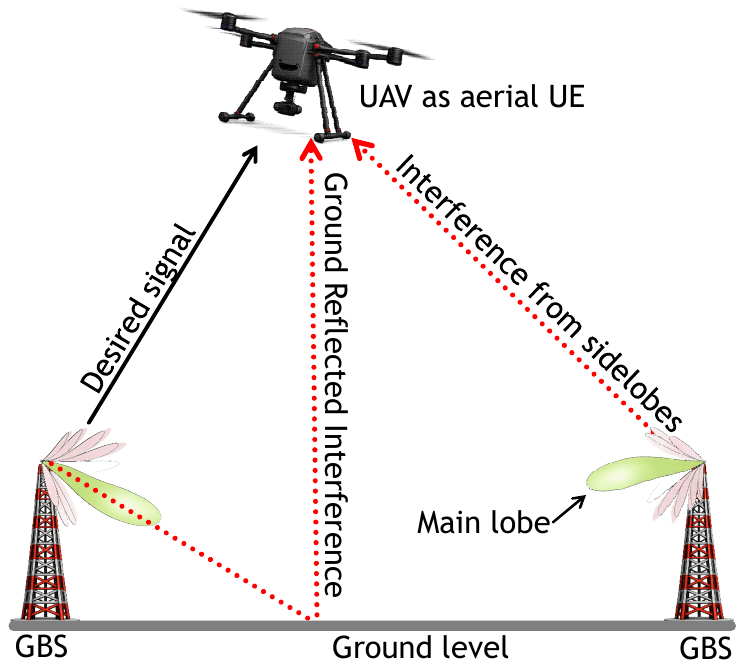}
   \caption{Illustration of altitude-dependent multi-cell interference affecting the reliability of EC in cellular-connected UAVs. 
   }
    \label{fig:framework}    
\end{figure}

Electronic conspicuity can be achieved through multiple technological approaches, including traditional aviation systems and emerging communication-based solutions~\cite{EC_nassif2024}. Among these, cellular networks have emerged as a promising enabler due to their pervasive infrastructure, wide-area coverage, and support for reliable communication links. However, as illustrated in Fig.~\ref{fig:framework}, cellular-connected UAVs operating at elevated altitudes experience unique propagation conditions that fundamentally differ from those of terrestrial users. Due to reduced blockage and increased line-of-sight (LOS) visibility, UAVs can simultaneously receive signals from multiple base stations (BSs), which improves coverage but also introduces strong inter-cell interference (ICI) in frequency reuse-1 cellular deployments. As a result, the communication link supporting EC becomes increasingly influenced by interference rather than noise, particularly at higher altitudes~\cite{hossen2026aerialboostercellenabledintercell}.
\begin{figure*}[t]
    \centering
    \includegraphics[width=1\linewidth,trim=0in 2.2in 0in 2.7in,clip]{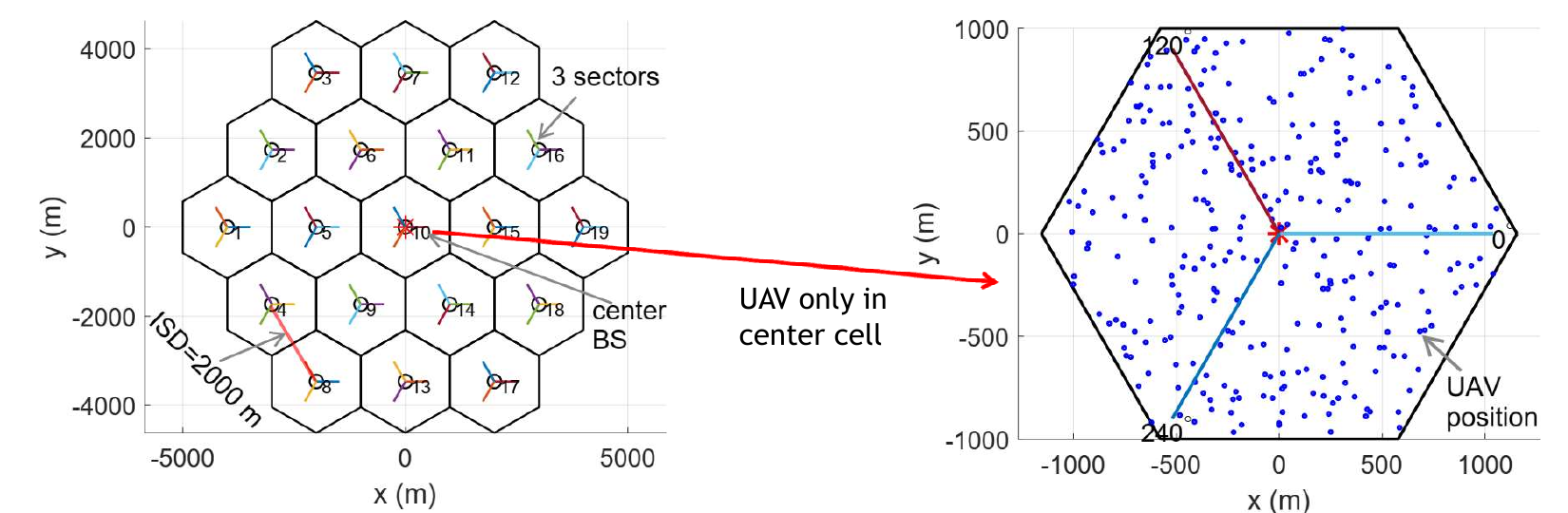}
    \caption{Multi-cell hexagonal wrap-around geometry with three-sector BSs for a representative inter-site distance (ISD) of 2000 m. UAV positions are considered within the central cell to evaluate altitude-dependent performance under multi-cell interference.}
    \label{fig:layout}
\end{figure*}
A key challenge in this setting is that conventional cellular deployments are optimized for ground users. BS antennas are typically downtilted, causing UAVs to be served through antenna sidelobes while also receiving comparable signals from multiple neighboring sectors. This leads to ambiguous serving-cell selection, fragmented association regions, and increased sensitivity to spatial variations. Consequently, UAV connectivity becomes less stable, and link quality can degrade despite strong received power. These effects directly impact the reliability of EC-related communication, where consistent and predictable connectivity is required.

Prior studies~\cite{3GPP36777,Zeng7470933, Maeng10416187} have primarily examined cellular-connected UAV performance using link-level metrics such as reference signal received power (RSRP), reference signal received quality (RSRQ), and signal-to-interference-and-noise ratio (SINR). While these metrics provide useful snapshots of link quality, they do not fully capture how altitude influences multi-cell interaction and association behavior across the network. Measurement-based studies~\cite{Maeng10416187, Raouf10200994} have further shown that aggregate received power tends to increase with altitude due to enhanced LOS conditions but also exhibits greater variability, indicating simultaneous exposure to multiple transmitting sources. 

This article provides a system-level perspective on how UAV altitude shapes the radio environment and impacts the reliability of EC in cellular-connected UAVs. Rather than relying only on link-level metrics, we examine spatial and network-level characteristics, including serving distance, association structure, and aggregate received power. These metrics provide an interpretable view of how UAV connectivity varies with altitude, network density, and propagation conditions. Our results show that increasing altitude strengthens multi-cell interaction and alters association behavior, leading to interference-dominated conditions at higher altitudes. These effects are further supported through measurement data from a helikite-based spectrum monitoring campaign and corresponding simulation results, which exhibit consistent altitude-dependent trends. The findings provide practical insights for designing altitude-aware and interference-aware cellular systems to support reliable UAV operation.

The main contributions of this work are summarized as follows:
\begin{itemize}
\item We provide a unified system-level characterization of altitude-driven multi-cell interaction in multi-cell networks, revealing how UAV connectivity becomes interference-limited at higher altitudes and impacts EC reliability.

\item We identify the role of antenna sidelobe-based reception in enabling simultaneous visibility of multiple co-channel sectors (COS), which leads to strong multi-cell interaction and fragmented association regions at higher altitudes.

\item We introduce serving distance as a new spatial metric to characterize association locality and spatial variability, showing how increasing altitude shifts connectivity from nearby to distant sectors.

\item We analyze the joint impact of UAV altitude, network density, and propagation environment, revealing a fundamental tradeoff between interference in dense deployments and path loss in sparse deployments.

\item We complement the system-level analysis with measurement-based validation using helikite spectrum monitoring data, showing similar altitude-dependent trends between real-world observations and simulations.
\end{itemize}

\section*{System Model}
\label{sec:system_model}
We consider a multi-cell cellular network supporting aerial users in the form of UAVs. The network consists of BS sites arranged in a hexagonal layout, as illustrated in Fig.~\ref{fig:layout}. To ensure boundary effects and consistent interference, wrap-around geometry is employed so that edge users experience interference equivalent to that in an infinite cellular layout~\cite{3gpp38901}. Each site is equipped with three sectorized antennas covering different azimuth directions. All sectors operate under a reuse-1 configuration, sharing the same frequency resources. The UAV operates within a central cell while being exposed to surrounding transmissions. Unlike~\cite{3gpp38901} terrestrial users, the UAV experiences a different propagation environment due to its elevated altitude, which significantly alters both signal reception and interference conditions. Here, we focus on how altitude, spatial location, and network density influence UAV association behavior and link quality in such multi-cell environments, which directly impacts the reliability of EC communication.
\begin{figure*}[t]
    \centering
    \includegraphics[width=0.92\linewidth]{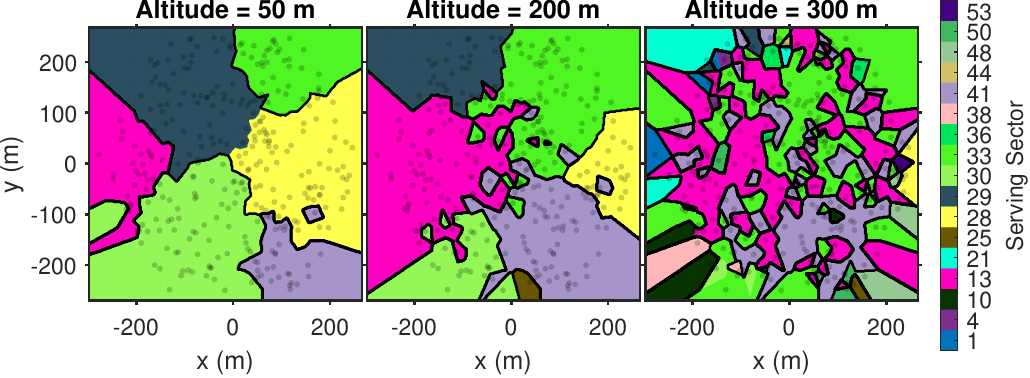}
    \caption{Serving-sector association maps for a rural scenario (ISD = 500 m) based on 300 sampled UAV positions (dots) at three altitudes. Colors indicate the most frequent serving sector with consistent mapping across altitude cases; background regions are obtained via nearest-neighbor interpolation for visualization.}
\label{fig:alt_sector}
\vspace{-.2in}
\end{figure*}

\textit{Propagation Characteristics:}
The simulation uses a 3.5 GHz carrier with a simplified environment-dependent path-loss model for urban macro (UMa) and rural macro (RMa) settings, including sector antenna gain, shadow fading, implementation loss, and altitude-dependent clutter attenuation. 
As the UAV altitude increases, the LOS conditions with multiple BSs also increase. While enhanced LOS improves the received signal strength from multiple directions, it simultaneously exposes the UAV to stronger interference from COS. At higher altitudes, interference from multiple sectors becomes dominant, which can degrade the reliability of EC communication.

\textit{Antenna Radiation and Sidelobe Effects:}
BS antennas are typically downtilted to optimize coverage for terrestrial users~\cite{Muruganathan_9696263, 3gpp38901}. 
Although sidelobes provide sufficient signal strength for connectivity, they also enable the UAV to receive comparable signals from multiple sectors. As shown in Fig.~\ref{fig:framework}, this results in simultaneous exposure to desired and interfering signals from multiple COS, increasing aggregate interference and reducing the dominance of any single serving BS.

\textit{Association and Serving Distance:}
The serving sector is selected using a hybrid association rule based on received signal strength and signal quality. Specifically, the UAV first identifies the sectors whose RSRP values lie within a small margin of the strongest received signal. If a single dominant sector exists, that sector is selected. When multiple candidate sectors provide comparable RSRP, the final association is determined by the sector with the best RSRQ. This reflects that signal strength alone may be insufficient for aerial users operating under strong multi-cell interference.

To characterize the resulting association behavior, we use the \textit{serving distance}, defined as the distance between the UAV and its serving BS. Serving distance provides a direct and physically interpretable measure of sector dominance and association locality. At low altitudes, the UAV is typically associated with nearby sectors, resulting in shorter serving distances. As altitude increases, multiple sectors become simultaneously visible through sidelobes, and the serving distance tends to increase, indicating reduced dominance of nearby cells and stronger multi-cell interaction. After association, the link is evaluated using standard cellular performance indicators, including RSRP, RSRQ, and SINR. These metrics are then used to quantify how altitude, propagation environment, and network density influence UAV multi-cell interaction behavior in reuse-1 cellular deployments.

\textit{UAV Sampling and Spatial Variability:}
To capture the spatial variability of UAV connectivity, multiple UAV positions are considered within the central cell at different altitudes. Rather than focusing on specific trajectories, the analysis evaluates representative UAV locations across the cell area. This approach enables characterization of altitude-dependent association behavior, spatial variations in serving distance, and exposure to multi-cell interference. The resulting analysis captures both average trends and distributional characteristics of UAV connectivity, providing insight into how multi-cell interaction behavior varies across space and altitude.

\textit{Performance Indicators:}
To analyze UAV multi-cell interaction behavior, we rely on standard cellular indicators and spatial association characteristics, including the following:
\begin{itemize}
    \item \textit{RSRP}, which determines received signal strength and initial association;
    \item \textit{RSRQ}, which reflects interference and network load conditions;
    \item \textit{SINR}, which determines link quality and achievable performance; and
    \item \textit{Serving distance}, which captures spatial association behavior and sector dominance.
\end{itemize}

These metrics capture signal strength, interference conditions, and spatial association behavior. By examining how these quantities evolve with altitude, UAV location, and network density, we analyze the interference dynamics in reuse-1 cellular networks, which directly influence the reliability of EC communication.

\section*{UAV Association Dynamics and Multi-cell Interaction}
\label{sec:results_discussion}
The following results quantify how altitude affects association structure, serving distance, and spatial variability in UAV connectivity.
\begin{figure*}
    \centering
    \includegraphics[width=.95\linewidth]{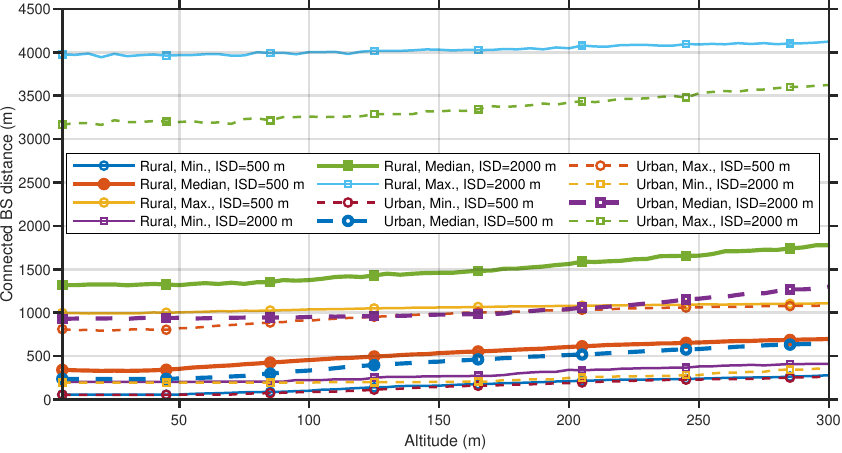}
    \caption{Connected BS distance versus UAV altitude for rural and urban scenarios. Minimum (Min.), median, and maximum (Max.) distances are shown for two different ISDs, with increasing spread at higher altitudes.}
\label{fig:serving_dis_alt}
\end{figure*}

\textit{Altitude-Driven Association and Interference Behavior: }Fig.~\ref{fig:alt_sector} illustrates the serving sector association maps at three representative altitudes. At low altitude (50 m), the UAV is primarily served by a small set of nearby sectors, resulting in smooth and stable association patterns. At an intermediate altitude (200 m), the association patterns begin to deform and fragment as the UAV becomes visible to more sectors, leading to more comparable received signal levels from multiple directions. At a higher altitude (300 m), this effect becomes more pronounced, resulting in strong multi-sector interactions and a loss of dominance by any single serving sector. Consequently, the association regions become fragmented and irregular, and small variations in UAV position lead to frequent changes in the serving sector. This reflects reduced association stability and highlights the impact of altitude on multi-cell interaction in aerial cellular networks. This increased association instability can negatively impact EC reliability, as variations in serving sector association may reduce link stability for EC operation.

\textit{Serving Distance with Altitude:}
Fig.~\ref{fig:serving_dis_alt} shows how the connected BS distance changes with UAV altitude for rural and urban scenarios at ISDs of 500 m and 2000 m. As the UAV altitude increases, the typical (median) serving distance gradually increases, i.e., it tends to connect to BSs that are farther away. At the same time, the gap between the minimum and maximum distances becomes larger, reflecting more variation across different UAV positions within the simulated area and less consistent association. A larger ISD results in overall higher serving distances due to wider spacing between BSs, while both rural and urban scenarios exhibit similar increasing trends with altitude, although their absolute distance levels differ. The increase in serving distance further contributes to EC reliability challenges by weakening signal dominance and increasing exposure to interference.
\begin{figure*}[t]
\centering
\includegraphics[width=0.31\linewidth]{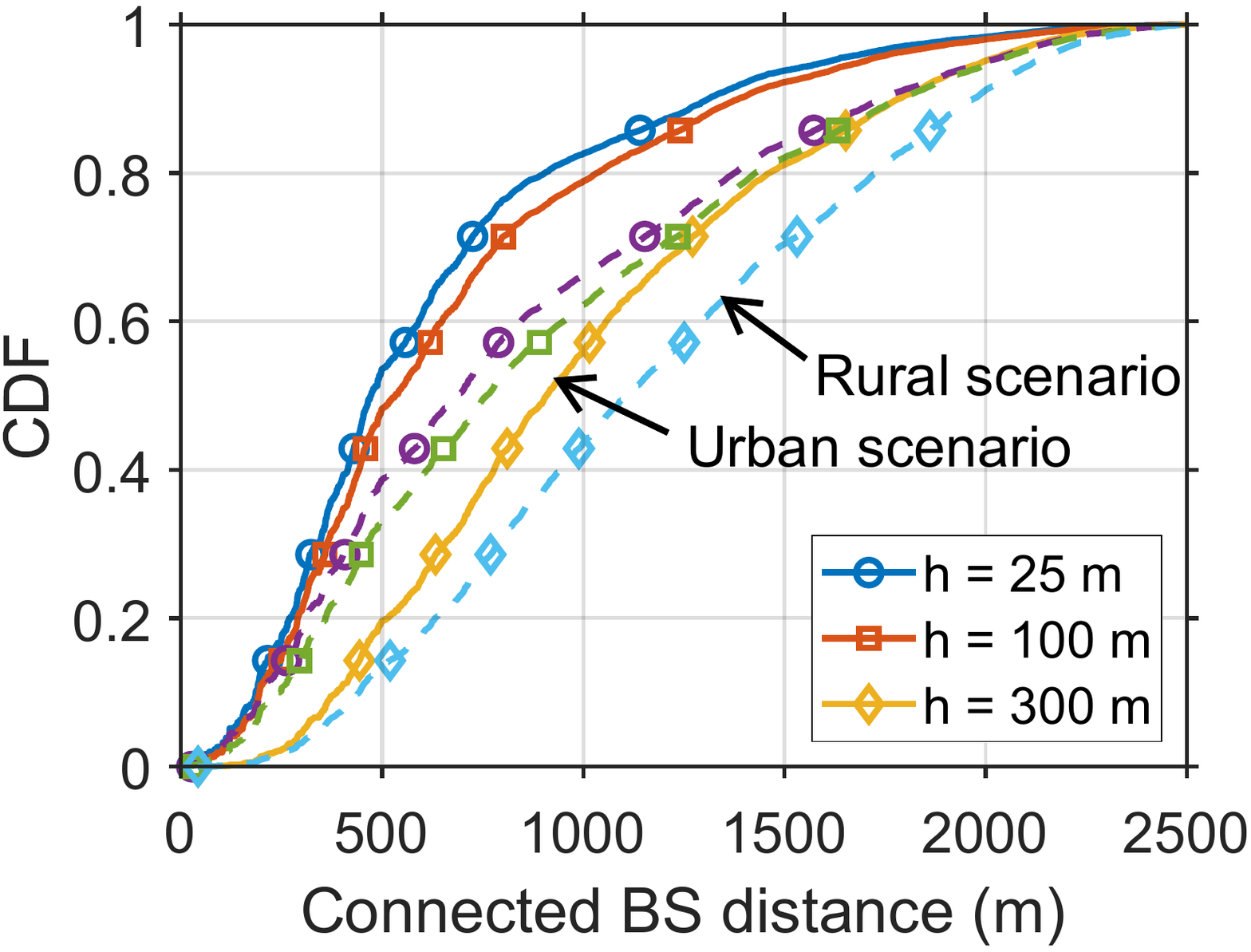}
\hfill
\includegraphics[width=0.31\linewidth]{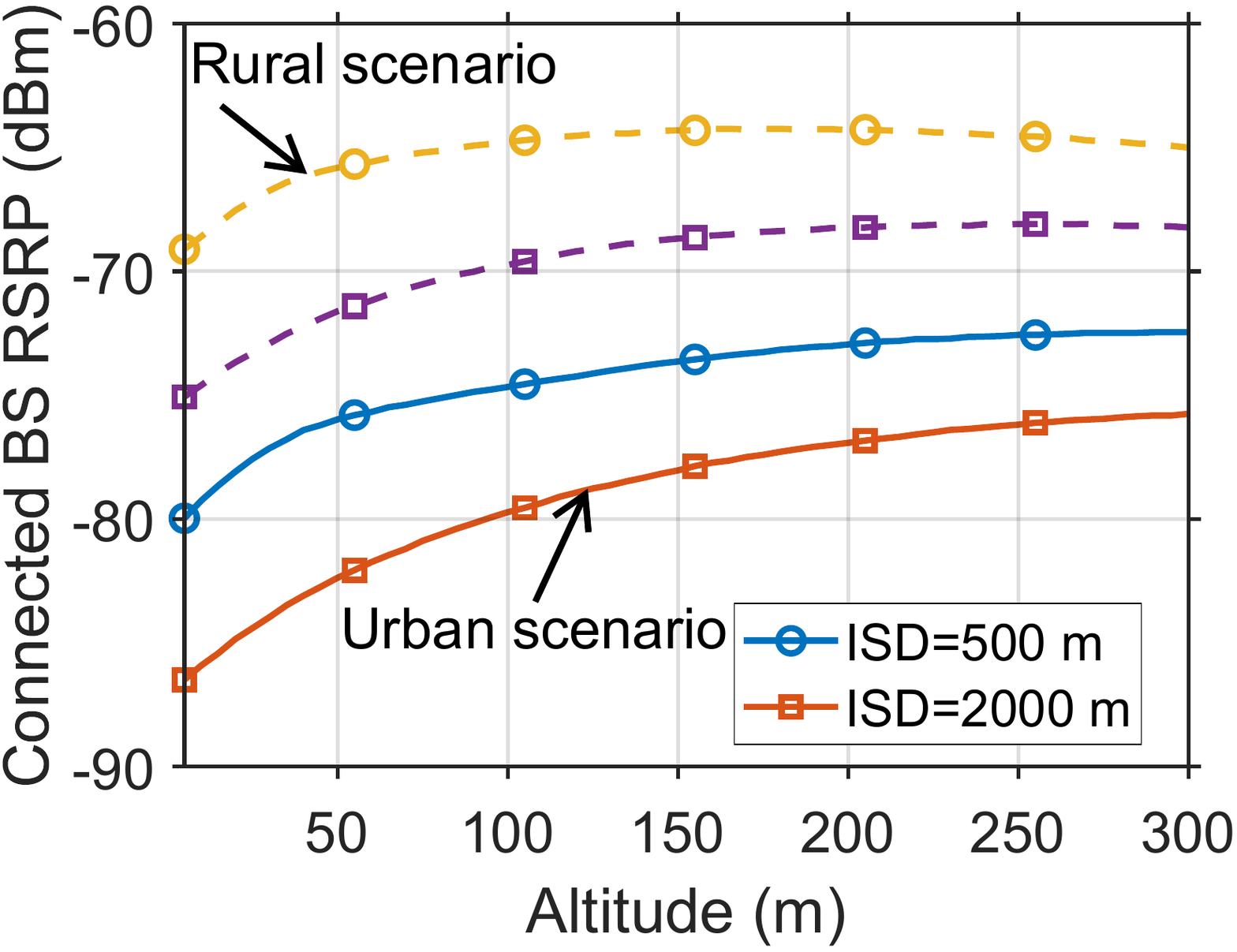}
\hfill
\includegraphics[width=0.33\linewidth]{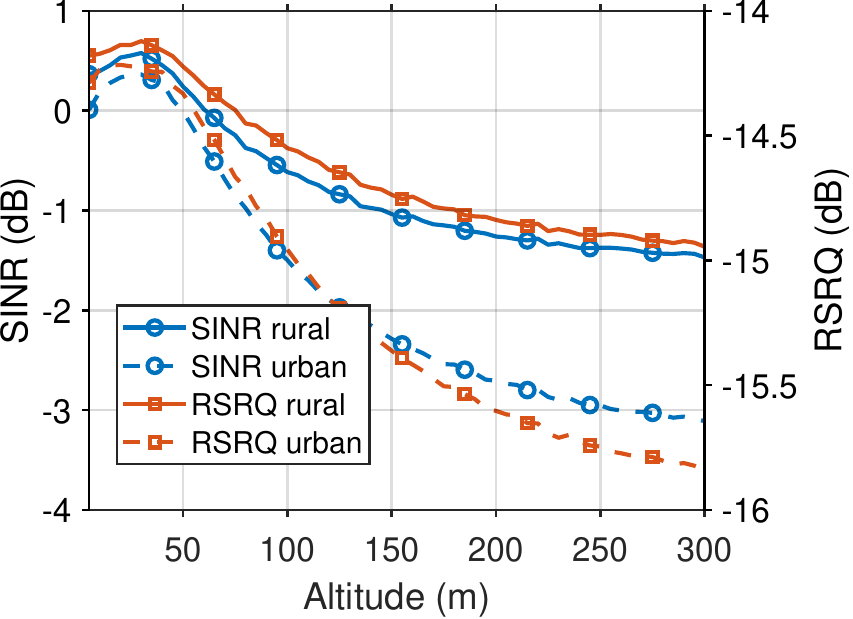}

\vspace{4pt}
\caption{Connected BS distance distribution (left), connected BS RSRP (middle), and SINR and RSRQ variation (right) with UAV altitude. The distance distribution shifts toward larger values at higher altitudes, indicating an increased likelihood of association with farther sectors. The RSRP reflects the impact of ISD and altitude on signal strength and tends to level off at higher altitudes, while SINR and RSRQ indicate degradation in link quality due to increased multi-cell interference.}
\label{fig:serv_dist_combined}
\end{figure*}

\begin{figure*}[t]
    \centering
    \includegraphics[width=0.28\textwidth]{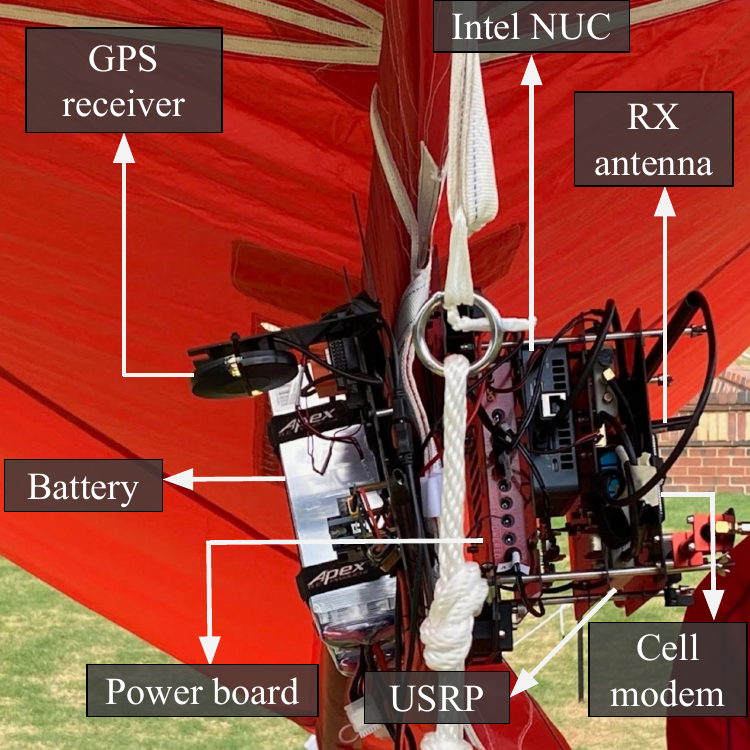}
    \hfill
    \includegraphics[width=0.2\textwidth]{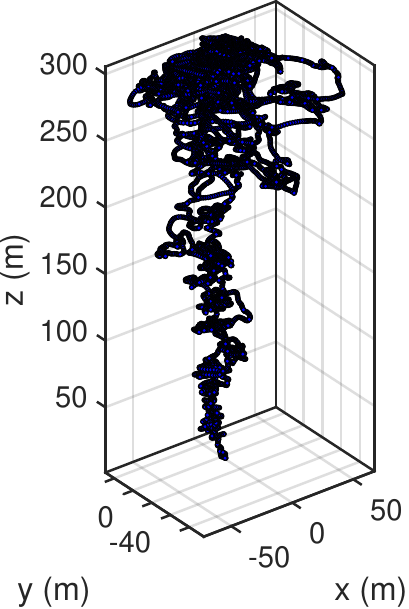}
    \hfill
    \includegraphics[width=0.38\textwidth]{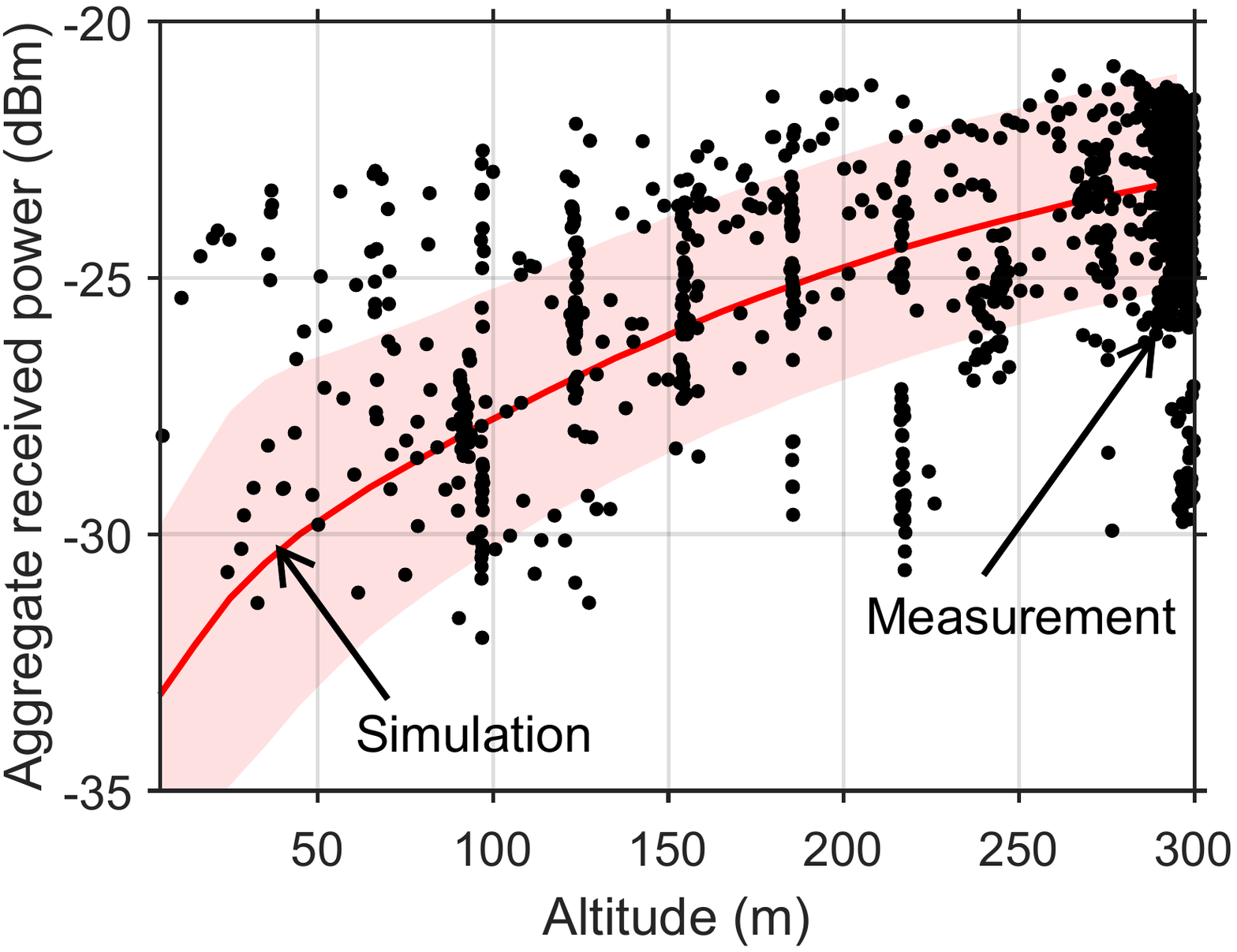}

\caption{Helikite-mounted measurement platform (left), the three-dimensional helikite trajectory during the measurement campaign (middle), and the measured and simulated aggregate received power versus altitude (right). The simulation shows the mean with one standard deviation, and the measurement data are shown as scatter points.}
\label{fig:helikite_measurement_comparison}
\end{figure*}

\textit{Serving Distance Distribution and Network Density Effects: }
While Fig.~\ref{fig:serving_dis_alt} captures average trends, it does not reveal how the association varies across different UAV locations. To capture this spatial variability, Fig.~\ref{fig:serv_dist_combined} presents the distribution of serving distance (left), along with the corresponding RSRP (middle), SINR, and RSRQ variations (right) with altitude. In both rural and urban scenarios, the serving distance distributions shift toward larger distances as altitude increases. This indicates that, for a significant fraction of UAV positions, the serving sector is no longer among the closest BSs. At low altitude (e.g., 25 m), most associations occur within short distances, as reflected by the steep rise of the cumulative distribution function (CDF). At higher altitudes (e.g., 300 m), the distribution becomes more gradual and extends to much larger distances, showing that associations are spread over a wider spatial range. This shift reflects a transition from localized association to broader interaction with multiple sectors across the network. Even when associated with a given BS, the UAV is influenced by other co-channel sectors, reducing the dominance of the serving link. 

This change in association behavior is reflected in link-level performance metrics. As the UAV ascends, the RSRP  improves with altitude and tends to level off at higher altitudes. Larger ISD values consistently result in lower RSRP because the serving BSs are farther away, leading to higher path loss. However, this relatively stable signal strength does not indicate better performance. As shown in Fig.~\ref{fig:serv_dist_combined}, the UAV is simultaneously exposed to stronger interference from multiple sectors at higher altitudes, leading to degradation in SINR and RSRQ. This highlights a fundamental tradeoff in network density, where dense deployments increase interference, while sparse deployments weaken the desired signal. Since EC depends on reliable communication links, this degradation can directly affect its reliability.

\textit{Measurement-Based Validation and Comparison: }
To complement the simulation-based analysis, we include a measurement example from the AERPAW helikite-based spectrum monitoring campaign conducted at the Lake Wheeler site in August 2024~\cite{raouf2024aerpaw, raouf2026curatedwirelessdatasetsaerial}, which is a rural environment. Fig.~\ref{fig:helikite_measurement_comparison} shows the helikite-mounted measurement platform on the left and the corresponding 3D trajectory during the campaign in the middle. The measurements are obtained via passive wideband spectrum sensing across the sub-6 GHz range, capturing aggregate received power from multiple cellular bands. 
These observations are consistent with prior spectrum monitoring studies, which report increasing aggregate received power with altitude due to enhanced LOS conditions and exposure to multiple transmitting sources~\cite{Raouf10200994}. In this experiment, the helikite ascends to approximately 300 m, remains at high altitude for several hours, and then descends, while a USRP-B205mini performs continuous wideband spectrum sensing across the sub-6 GHz cellular bands. The helikite measurement captures aggregate received power from surrounding cellular transmissions without knowledge of transmitter locations or configurations. In contrast, the simulation assumes a structured multi-cell deployment with known BS locations and standardized propagation models. Despite these differences, both approaches exhibit similar altitude-dependent trends.
Furthermore, the helikite follows a single trajectory with limited horizontal variation, whereas the simulation evaluates multiple UAV positions distributed across the cell area at each altitude. This enables the simulation to capture spatial variability in association and interference as reflected in the spread on the right in Fig.~\ref{fig:helikite_measurement_comparison}. In both cases, received power generally increases with altitude due to enhanced LOS conditions and increased exposure to multiple transmitting sources. This reflects realistic aggregate interference conditions experienced by aerial users in operational environments and suggests that altitude-dependent interference can directly impact the reliability of EC in practical deployments.

\section*{Design Implications for UAV-Enabled Cellular Networks}
The observed changes in association structure and serving distance provide direct insight into how UAV operation should be adapted in practical deployments. The results show that UAV altitude fundamentally reshapes association behavior and interference conditions in cellular networks. These observations lead to several important design insights.

\textit{Altitude-Aware Operation: }As shown by the increase in serving distance and multi-sector visibility, higher altitude improves visibility but also increases exposure to multiple co-channel sectors. UAV operation should therefore balance coverage benefits with the resulting increase in interference. Selecting moderate altitude levels can help maintain more stable connectivity for reliable EC operation.

\textit{Interference-Aware Network Design: }
At higher altitudes, UAV connectivity becomes interference-limited. Network design and resource management strategies should explicitly account for aggregate interference from multiple sectors, rather than relying on terrestrial-centric assumptions for reliable EC communication. As observed in Fig.~\ref{fig:serv_dist_combined}, SINR degradation at higher altitudes reinforces the need for interference-aware design strategies. Techniques such as inter-cell interference coordination (ICIC) and adaptive power control can help mitigate these effects and improve the reliability of EC communication in such scenarios.

\textit{Robust Association Strategies: } Association based solely on signal strength is insufficient for aerial users. At higher altitudes, multiple sectors provide comparable signal levels, leading to unstable connections. Incorporating signal quality indicators (e.g., RSRQ) into association decisions can improve stability and reduce unnecessary switching between sectors.

\textit{Deployment Density Considerations: } Network density plays a key role in UAV performance. Dense deployments increase the number of visible sectors and strengthen interference, while sparse deployments increase serving distance and path loss. Effective deployment strategies should balance these effects to support reliable UAV connectivity.

\textit{Environment-Aware Adaptation:} Propagation conditions vary across environments. Rural scenarios expose UAVs to stronger multi-cell visibility, while urban environments may partially limit interference due to blockage. UAV operation and network configuration should adapt to these environment-specific characteristics.

\section*{Conclusion}
This article presented a system-level analysis of UAV connectivity in multi-cell cellular networks, focusing on how altitude influences association behavior and interference conditions. The results show that increasing UAV altitude shifts connectivity from a signal strength–dominated regime to an interference-limited one. With enhanced visibility and sidelobe reception, UAVs become exposed to multiple co-channel sectors, leading to fragmented association, longer serving distances, and increased spatial variability. As a result, UAVs are often associated with non-nearest sectors, indicating a loss of locality in cell selection and greater sensitivity to multi-cell interference.

The analysis also shows that network density and propagation environment significantly influence these behaviors. Dense deployments increase the number of visible sectors and strengthen aggregate interference, while sparse deployments increase serving distance and path loss. Differences between rural and urban scenarios further highlight the role of propagation conditions in shaping interference and association patterns. These findings demonstrate that reliable UAV operation cannot be characterized by signal strength alone. Instead, it requires careful consideration of multi-cell interference, spatial association behavior, and altitude-dependent propagation effects. The consistency between simulation and measurement results reinforces that altitude-driven multi-cell visibility governs the aerial radio environment in practical deployments and directly impacts EC reliability.


\ifCLASSOPTIONcaptionsoff
  \newpage
\fi



%

\bibliographystyle{IEEEtran}
\bibliography{ref}

\end{document}